# Brain Emotional Learning-based Prediction Model
For the Prediction of Geomagnetic Storms

Mahboobeh Parsapoor

*Abstract—* **This study suggests a new data-driven model for the prediction of geomagnetic storm. The model which is an instance of Brain Emotional Learning Inspired Models (BELIMs), is known as the Brain Emotional Learning-based Prediction Model (BELPM). BELPM consists of four main subsystems; the connection between these subsystems has been mimicked by the corresponding regions of the emotional system. The functions of these subsystems are explained using adaptive networks. The learning algorithm of BELPM is defined using the steepest descent (SD) and the least square estimator (LSE). BELPM is employed to predict geomagnetic storms using two geomagnetic indices, Auroral Electrojet (AE) Index and Disturbance Time (Dst) Index. To evaluate the performance of BELPM, the obtained results have been compared with ANFIS, WKNN and other instances of BELIMs. The results verify that BELPM has the capability to achieve a reasonable accuracy for both the short-term and the long-term geomagnetic storms prediction.**

## I. Introduction

The geomagnetic storm originates from the solar wind which disturbs the Earth's magnetosphere and has a direct association with the solar cycle [1]. Geomagnetic storms have caused harmful damage to radio communication, the orbit of satellite, power grids, etc. Thus, the prediction of geomagnetic storms is very important to prevent these harmful effects. Two important indices to predict geomagnetic storms are Auroral Electrojet (AE) index and Disturbance Time (Dst) index [2]-[15].

Recently, taking inspiration from the mammalian emotional systems to develop emotion-based decision-making, emotion-based controllers and emotion-based machine learning approaches have received a lot of attention [16]. Among them, emotion-based machine-learning approaches and their prediction applications have been more favorable than others. They have showed a high generalization capability to model nonlinear behavior of chaotic time series. The fundamental model of most of emotion-based machine-learning approaches is the amygdala-orbitofrontal system which is a computational model of emotional learning. This model has a simple structure and imitates the interaction between some parts of the emotional system (e.g., the amygdala, thalamus, sensory cortex and orbitofrontal) and formulates the emotional response using mathematical equations [17].

This paper suggests another framework for emotion-based machine learning approaches being applied as the prediction models. The performance of this framework is evaluated by applying it to predict geomagnetic storms. The main contribution of this paper is to present a useful prediction model for both the short-term and long-term prediction in space weather applications.

The rest of the paper is organized as follows: Section II gives a review of bio-inspired prediction models. Section III describes the general structure of BELIMs. Section IV compares well-known data-driven model and BELPM. In Section V related studies in the prediction of geomagnetic storms are reviewed and BELPM is tested to predict the indices of geomagnetic storms. Finally, conclusions about the performance of BELPM and the further improvements to the model are discussed in Section VI.

## II. Bio-inspired prediction models

Developing bio-inspired prediction models is one of the hot research topics in the computational intelligence community. Earlier studies are related to mimic the mammalian nervous system to develop artificial neural networks that have shown flexibility, robustness and generalization capability to predict the nonlinear and chaotic behavior of the complex system [18]. Different types of neural networks, multilayer perceptron, and radial basis function and recurrent neural networks have been proposed for the prediction applications [18]-[22]. Another well-known bio-inspired prediction model is the neuro-fuzzy model, which combines adaptive network and fuzzy logic. It has been proven that the neuro-fuzzy model has the capability to predict chaotic time series with an arbitrary accuracy [23].

Hierarchical Temporal Memories (HTM) is another prediction model that has been developed by studying the human neurocortex. HTM has a hierarchical structure which mimics the structure of the neurocortex, which 'is a large sheet of neural tissue about 2mm thick' [24]. HTM has been successfully employed to predict financial time series [24]-[25]. Recently, emotion-based machine-learning approaches have been proposed for chaotic time series prediction. They have been developed by the classic conditioning aspect of emotional processing and can be classified on the basis of their fundamental frameworks [16].

## III. A Brain Emotional Learning-based Inspired Model

Emotion and the emotional processing have been active research topics for neuroscientists and psychologists and a lot of effort has been made to analyze emotional behavior and describe emotion on the basis of different hypotheses, e.g., psychological, neurobiological, philosophy, and learning hypothesis. These hypotheses have contributed to the present computational models of emotion that are computer-based models of emotional processing [16][17]. The computational model has been reviewed in [16].

One simple computational model is the amygdala-orbitofrontal system that has been the fundamental basis of several emotion-based machine learning models. It was developed on the basis of the internal structure of the emotional system. The amygdala-orbitofrontal system consists of four parts which interact with each other to form the association between the conditioned and the unconditioned stimuli (see Fig. 1) [17]. In this model, the orbitofrontal and amygdala are represented by several nodes with linear functions. The output vector of the amygdala and the orbitofrontal cortex are referred to as $A$ and $O$, respectively. The output of the model is represented as $E$ and it is formulated as equation (1).

$$\mathbf{E} = \sum_i A_i - \sum_i O_i \quad (1)$$

Here $A_i$ and $O_i$ are the output of $i^{th}$ node of the amygdala and the orbitofrontal part. The updating rules of the model are based on $A$, $O$ and the reinforcement signal $REW$. The updating rules are formalized as equations (2) and (3) and are utilized to adjust the weights $\mathbf{V}$ and $\mathbf{W}$ are associated with nodes of the amygdala and the orbitofrontal part, respectively [17]. Here $s_i$ is the input stimulus for the weight $i^{th}$ node of the amygdala and the orbitofrontal part.

$$\Delta V_i = \alpha (S_i \times \max(0, \mathbf{REW} - \sum_i A_i)) \quad (2)$$

$$\Delta W_i = \beta (S_i \times (\sum_i O_i - \mathbf{REW})) \quad (3)$$

The basic amygdala-orbitofrontal model has a simple structure and can be used as a foundation for emotion-based machine learning approaches [17].

### A. Structural Aspect of the Brain Emotional Learning-based Model

The structure of BELPM is based on the general structure of BELIMs that has been depicted in Fig. 2. This structure is copied by the connection of those parts of the brain that are responsible to the emotional learning process. As mentioned, the structure of the model consists of four main parts (See Fig. 2): TH, CX, AMYG and ORBI which refer to the THalamous, sensory CorteX, AMYGdala, and ORBItofrontal cortex, respectively. These parts have important roles in emotional learning. Certainly, the emotional system's regions are very complex and this structure has not mimicked all their connections in detail. The suggested structure is also the foundation of the Brain Emotional Learning-based Fuzzy Inference System (BELFIS), the Brain Emotional Learning-based Recurrent Fuzzy System (BELRFS) and the Emotional Learning Inspired Ensample Classifier (ELiEC) [27]-[29].

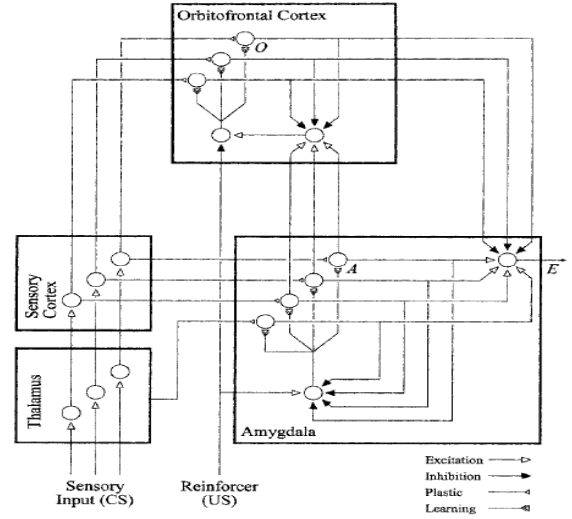

Fig. 1. The Amygdala-orbitofrontal [17].

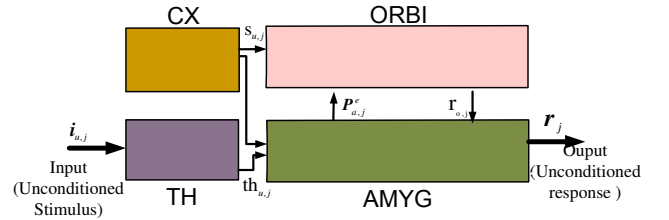

Fig. 2. The structure of BELPM. This structure is the general structure of BELIMs.

The following steps explain the connection and input and output of each part of BELPM when it receives a conditioned stimulus as $i_{u,j}$. Note, two subscripts $u$ and $c$ have been utilized to distinguish the training data set and the test data set that are defined as $I_c = \{\mathbf{i}_{c,j}, r_{c,j}\}_{j=1}^{N_c}$ and $I_u = \{\mathbf{i}_{u,j}, r_{u,j}\}_{j=1}^{N_u}$, with $N_c$ and $N_u$ data samples, respectively. 1) The TH has a connection with the AMYG and the CX; thus it receives the input vector, $i_{u,j}$ and sends $th_{u,j}^{AGG}$ and $th_{u,j}^{Max\_Min}$ to CX and AMYG, respectively. 2) CX provides $\mathbf{s}_{u,j}$ and distributes it between the AMYG and the ORBI. 3) The AMYG receives two inputs, $th_{u,j}^{Max\_Min}$ and $\mathbf{s}_{u,j}$, which originated from the TH and the CX. This part provides $r_{a,j}$ and $p_{a,j}^e$, the primary response and the expected punishment, respectively. There is a bidirectional connection between this part and the ORBI (see Fig. 2). To imitate the amygdala region and its components, AMYG is divided into two components: BL (corresponds to the set of the Basal and Lateral) and CM (corresponds to the set of accessory basal and CentroMedial parts). 4) The ORBI receives $s_{u,j}$ and $p^e_{a,j}$ and provides $r_{o,j}$ and send it to AMYG. 5) The final output, $r_j$, is provided using the primary and secondary outputs. ORBI is also divided into two parts: MO (Medial of Orbitofrontal) and LO (Lateral of Orbitofrontal). It imitates the roles of the orbitofrontal to form a stimulus-reinforcement association, evaluate reinforcement, and provide an output. It starts performing its functions after receiving the expected punishment, $p^e_{a,j}$, from CM, which means that the BL of AMYG must have fulfilled its functions.

### B. Functional Aspect of the Brain Emotional Learning-based Model

The functional aspect of BELPM can be explained using a simple adaptive network that has been depicted in Fig. 3. This adaptive network has been divided into four layers. In the following, the function, input and the output of each layer has been explained.

The first layer consists of $k_a$ adaptive or square nodes with K(.) function (kernel function). Each node has an input that is an entity from the $\mathbf{d}_{a\min} = \{d_{a\min,j}\}_{j=1}^{k_a}$ (which is a set of $k_a$ minimum distances of $\mathbf{d}_a = \{d_{a,j}\}_{j=1}^{N_u}$). The distances can be calculated as Euclidean distances between a new input as $\mathbf{i}_{c,j}$ and the training data as $\{\mathbf{i}_{u,j}\}_{j=1}^{N_u}$. The output vector of the first layer that is calculated using (4) is $\mathbf{n}^1_a$. Here, the input to the $m^{th}$ node is $d_{a\min,m}$.

$$n^1_{a,m} = K(d_{a\min,m}) \qquad (4)$$

In general, the kernel function for the $m^{th}$ node can be one of the functions that have been defined as (5), (6), and (7). The input and the parameter of K(.) of $m^{th}$ node can be determined using $d_m$ and $b_m$. The subscript a is used to distinguish the adaptive network of BL of AMYG andits related parameters ($\mathbf{d}_{a\min}$ and $\mathbf{b}_a$).

$$K(d_m) = \exp(-d_m b_m) \qquad (5)$$

$$K(d_m) = \frac{1}{(1+(d_m b_m)^2)} \qquad (6)$$

$$K(d_m) = \frac{\max(\mathbf{d}) - (d_m - \min(\mathbf{d}))}{\max(\mathbf{d})} \qquad (7)$$

The second layer is a normalized layer and has $k_a$ nodes (fixed or circle), which are labeled N to calculate the normalized value of $\mathbf{n}^1_a$ as $\bar{\mathbf{n}}^1_a$ using (8).

$$n^2_{a,m} = \frac{(n^1_{a,m})}{\sum_{m=1}^{k_a} n^1_{a,m}} \qquad (8)$$

The third layer has $k_a$ circle nodes with functions given in (9). This layer has two input vectors, $\bar{\mathbf{n}}^1_a$ and $\mathbf{r}_{ua}$; the latter is a vector that is extracted from $\mathbf{r}_u = [r_{u,1}, r_{u,2}, ..., r_{u,N_u}]$ and is related to the target outputs of the $k_a$ samples of $\{\mathbf{i}_{u,j}\}_{j=1}^{N_u}$ that have minimum distances with the new input $\mathbf{i}_{c,j}$.

$$n^3_{a,m} = \frac{(n^1_{a,m})}{\sum_{m=1}^{k_a} n^1_{a,m}} \times r_{ua,m} \qquad (9)$$

The fourth layer has a single node (circle) that calculates the summation of its input vectors, $\mathbf{n}^3_a$, to produce $r_a$.

The above explanation has illustrated the function of the simple adaptive network in BELPM. As mentioned, in BELPM, AMYG consists of two parts: BL and CM; while ORBI is divided into two parts: MO and LO. Figure 4 depicts the internal parts of BELPM and the connections between them during the first learning phase. Receiving $th^{AGG}_{u,j}$ and $th^{Max\_Min}_{u,j}$, BL provides $\mathbf{r}_{a,j}$ and sends to CM. During learning phase, BL also memorizes $r_j$ the corresponding target of the input vector, $i_{u,j}$ and sends it to CM which calculates the expected punishment $p^e_{a,j}$ and the punishment $p_{a,j}$. The former is sent to MO; while the latter is sent to BL. Mo provides the secondary response, $r_{o,j}$, and sends it to LO that is responsible to provide $p_{o,j}$, the punishment signal. Note that the updating rules are defined using the punishment signals. The connections between the adaptive network of BL of AMYG, the adaptive network of MO of ORBI and CM of AMYG have been depicted in Fig. 5. Note that BELPM has two phases: first learning phase and second learning phase; this figure describes the connection between BL, MO and CM during the first learning phase. It also shows how the functions of ORBI and AMYG can be explained adapting this simple adaptive network.

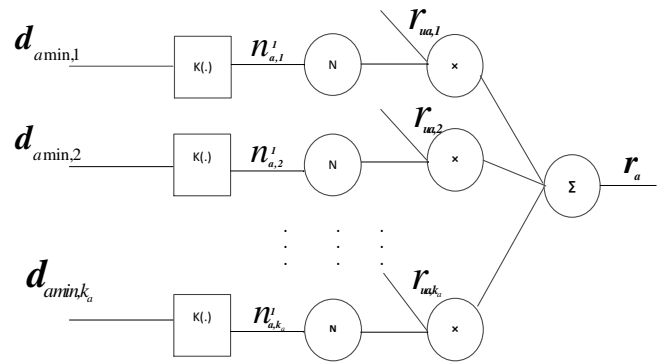

Fig. 3. A simple adaptive network.

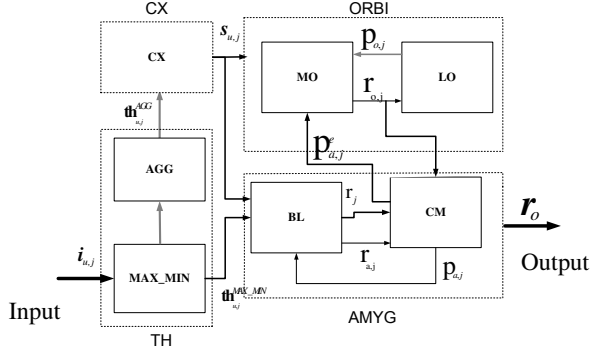

Fig. 4. The architecture of BELPM showing the structure of each part and its connection to other parts. An input from training set, unconditioned stimulus, enters the BELPM.

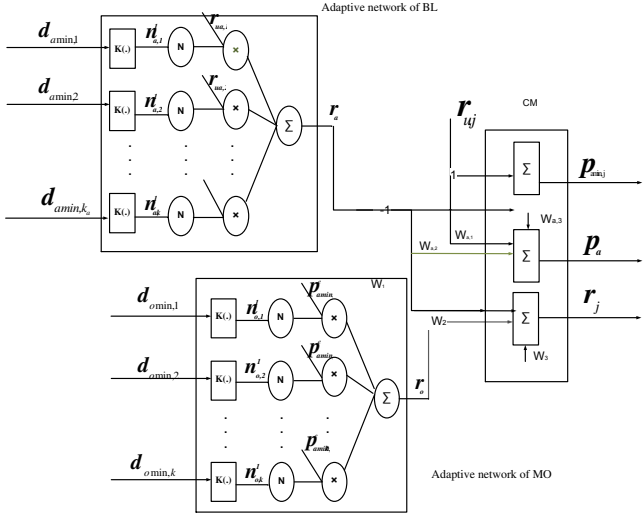

Fig. 5. Connection between the adaptive networks.

The function of BELPM was explained in detail in [30]. This paper briefly explains how each part is involved in providing the final output during the first learning phase. 1) The input vector $\mathbf{i}_{u,j}$ is fed to TH that consists of two subparts: MAX_MIN and AGG. MAX_MIN, which is a modular neural network and provides an output that is referred to as $\mathbf{th}_{u,j}^{\text{Max\_Min}}$. It calculates using (10) and sends it to both AGG and AMYG. Another part of TH, which is named the AGG, can also be described as a neural network with $R+2$ linear neurons; here R is the dimension of the input vector. The output of AGG, $\mathbf{th}_{u,j}^{\text{AGG}}$, which is calculated as (11), is equal to $\mathbf{i}_{u,j}$ and is fed to CX.

$$\mathbf{th}_{c,j}^{\text{Max-Min}} = [\text{Max}(\mathbf{i}_{c,j}), \text{Min}(\mathbf{i}_{c,j})] \qquad (10)$$

$$\mathbf{th}_{c,j}^{\text{AGG}} = \mathbf{i}_{c,j} \qquad (11)$$

2) $\mathbf{th}_{u,j}^{\text{AGG}}$ is sent to CX that provides $\mathbf{s}_{u,j}$ and distributes it between AMYG and ORBI. It should be noted that $\mathbf{i}_{u,j}$ and $\mathbf{s}_{u,j}$ have the same entity. However, they have originated from different parts.

3) Both $\mathbf{s}_{u,j}$ and $\mathbf{th}_{u,j}^{\text{Max\_Min}}$ are sent to AMYG. As mentioned earlier, AMYG is divided into two parts: BL and CM and provides the primary and final responses. The function of BL can be explained using the adaptive network. CM is responsible for providing the final output. It has inputs from BL and MO, and performs different functions during the first leaning phase and the second learning phase of the BELPM. Here, we focus on the function of CM during the first learning phase. In this phase, CM has three summation nodes. The first node and the second node receive $r_{u,j}$ and $r_{a,j}$, which that are the corresponding target of $\mathbf{i}_{u,j}$ and the primary responses of BL, respectivly. The third node receives $r_{a,j}$ and $r_{o,j}$, which are the primary responses of BL and MO. The function of the third node, second node and the first node are calculated as (12), (13)

$$r_j = w_1 r_{a,j} + w_2 r_{o,j} + w_3 \qquad (12)$$

$$p_{a,i} = w_{a,1} r_{u,i} + w_{a,2} r_{a,j} + w_{a,:} \qquad (13)$$

4) The expected reinforcements, $\mathbf{p}_a^e$, and $\mathbf{s}_{u,j}$, are sent to ORBI which is connected to CX and AMYG. The function of MO of ORBI can be explained using an adaptive network that has been described in Fig. 4.

LO evaluates the output of MO, generates $p_{o,j}$ as reinforcement (punishment), and sends it to MO. It has one node (square) with a summation function given in (14).

$$p_{o,j} = w_{o,1} r_{o,j} + w_{o,2} \qquad (14)$$

*C. Learning Aspect of Brain Emotional Learning-based Model*

To adjust the linear and nonlinear learning parameters, a hybrid learning algorithm that consists of the steepest descent (SD) and the least-squares estimator (LSE) is used. The SD updates the nonlinear parameters in a gradient related direction to minimize the loss functions, which are defined based on reinforcement signals $p_{a,j}$, $p_{a,j}^e$ and the outputs of the adaptive networks. The LSE is applied to update the linear parameters. Note the hybrid learning algorithm is independently applied to update the parameters of each adaptive network. The learning algorithm has been explained in detail in [30].

IV. A COMPARISON BETWEEN BELPM AND OTHER DATA-DRIVEN MODELS

BELPM differs from the previously proposed data-driven models in terms of prediction accuracy, structural simplicity and generalization capability. In the following, the differences between BELPM and other well-known data-driven models have been explained.

*1) Radial Bias Function (RBF)* differs from BELPM in terms of the underlying structure, inputs of the neurons, connection between neurons and the number of learning parameters and learning algorithms.

*2) Generalization Regression Neural Network (GRNN)* [18] differs from BELPM in its number of neurons (i.e., the number of neurons of GRNN are equal to the size of the training samples). Moreover, GRNN has no learning algorithm to optimize its performance and increase its generalization capability.

*3) Adaptive Neuro Fuzzy Inference System (ANFIS)* and BELPM are not similar because of different structures, functions and some aspects of learning algorithms. Due to the learning algorithm and the large number of learning parameters (linear and nonlinear) that are spread through the layers, ANFIS has the capability to obtain very accurate results for complex applications [19][23]. However, its learning algorithm has a significant effect on its computational complexity and it also causes over-fitting problems. The curse of dimensionality is another issue of ANFIS and increases the computational time of ANFIS for high-dimension application. Although the number of learning parameters of BELPM is not dependent on the dimension of input data, as mentioned before, BELPM uses Wk-NN; consequently, the computational time of BELPM only depends on the number of neighbors. To decrease its time complexity in high-dimension cases, we can choose a small number of neighbors for the BELPM.

*4) Local Linear Neuro Fuzzy Models (LLNF)* and BELPM can both be considered as types of "local modeling" [18][19] algorithms. They both combine an optimization-learning algorithm and LSE to train the learning parameters. However, LLNF uses the Local Linear Model Tree (LoLiMoT) algorithm, instead of the Wk-NN method of BELPM. The number of learning parameters of LoLiMoT has a linear relationship with the dimension of input samples and number of epochs; thus, its computational complexity has no exponential growth for high-dimension applications.

*5) Modular neural network* is a combination of several modules with different inputs [18] without any connection with others. There is no algorithm to update the learning parameters of the modules.

*6) Hybrid structures* that are defined in [18], differ from BELPM in receiving the input data. The sub modules of a hybrid structure can also be designed in parallel or series.

## V. Prediction Geomagnetic Storms

Prediction of geomagnetic storms is essential to prevent damage to any satellites' orbit and ground-based communications. So far, statistical methods as well as data driven models e.g., linear input-output techniques or linear prediction filtering neural network, neurofuzzy, have been examined to predict geomagnetic storms [10]-[12]. Most of these studies have utilized prediction models to predict two well-known indices: Disturbance Storm Time and Auroral Electrojet [1]-[15].

The disturbance storm time, Dst, is one of popular time series for examining statistical models and data-driven models and has been defined by Bruce Tsurutani. It is a measurement to count 'the number of solar charged particles that enter the Earth's magnetic field' [12]. Dst could be utilized to measure the intensity of geomagnetic storms and it has been recorded by several space centers such as World Data Center for Geomagnetism, Kyoto.

In [31] earlier studies related to use Dst to predict geomagnetic storms have been reviewed. In addition, a Neural network- based prediction model has been suggested to predict the minimum values of Dst during the recovery phase of geomagnetic storms. The model has been successfully examined to predict geomagnetic storms of 1980 and 1989. In [32], a recurrent neural network has been introduced to predict one hour step of Dst from 2001. This study also showed that combining principle components and NN causes a significant increase in prediction performance. The damage and harmful effects of geomagnetic storms have been reviewed in [9] where the variation of embedding dimension to analyze the chaotic Dst time series has been studied and tested for two super storms: 13 March 1989 and 11 January 1997. In [7] a combination of Singular Spectrum Analysis, SSA and locally linear neuro-fuzzy model have been proposed as useful methodologies to long term prediction of Dst time series. Specifically, this method has been examined to predict ten step ahead of extracted Dst time series between 1988 and 1990.

During this time, the geomagnetic storm damaged Quebec's power grid and caused a blackout in Quebec [8]. A nice review of Dst prediction models and the benefits of prediction Dst have been illustrated in [15]. Moreover, this study has proposed a long term prediction model that is called Anemomils. It has been tested for three geomagnetic storms, 2001, 2005 and 2012.

The prediction of future values of the AE index is also useful to forecast geomagnetic storms and sub storms. Auroral Electrojet, AE, has been proposed as a global quantitative index to characterize the magnetosphere's geomagnetic activities. AE has been defined to measure auroral zone magnetic activity by Sugiura and Davis [11]. The studies related to predict AE time series have been started in 1971. Various types of neural network, linear filter and nonlinear filter such as nonlinear moving average MA filter and nonlinear auto regressive moving average have been developed to predict AE time series [14]. In [15] a real time learning model has been proposed to predict AE and Dst . This paper also reviewed the prediction algorithms in space weather applications. In [33], a locally linear neuro-fuzzy model tree algorithm has been used to predict the AE index. In [34], an emotion-based machine- learning approach called BEL has been used to predict the AE index.

To provide a careful comparison with other methods, we used various data sets with different initialized points and sizes of training samples. This paper utilizes two error measures: normalized mean square error (NMSE) and mean square error (MSE), as given in (16) and (17), to assess the performance of the prediction models and provide results comparable with other studies. The correlation coefficient that is calculated as (18) is also utilized to compare the obtained results of the other studies.

$$\text{NMSE} = \frac{\sum_{j=1}^{N}(y_j - \hat{y}_j)^2}{\sum_{j=1}^{N}(y_j - \overline{y}_j)^2} \quad (16)$$

$$\text{MSE} = \frac{1}{N}\sum_{j=1}^{N}(y_j - \hat{y}_j)^2 \quad (17)$$

$$\rho_{y,\hat{y}} = \frac{\text{Cov}(y, \hat{y})}{\sigma_y \sigma_{\hat{y}}} \quad (18)$$

Where $\hat{y}$ and $y$ refer to the predicted values and desired targets, respectively. The parameter $\overline{y}$ is the average of the desired targets.

### A. Prediction AE Index

In this subsection, BELPM is tested for the AE index; the fractal dimension and the kolmogorove entropy of AE time series are 3.5 and 0.2 respectively [33]. It shows that the short-term state of AE time series is predictable [33] and the accurate prediction of the long-term state of the AE is almost impossible. In light of this fact, several data driven methods such as LoLiMoT [33], BEl, ANFIS, MLP [34], BELRFS and BELFIS [16] have been utilized to predict the AE time series.

As the first experiment of this subsection, BELPM has been employed for a long-term prediction of AE time series. Here, the AE time series is stated as the mean daily observations of geomagnetic storms from 1977 to 1987; the training samples are selected from 1977 to 1986 and the AE values of 1987 are used as test data. In this case study, the embedded dimension is selected as three. Table I compares the obtained NMSE of BELPM to the NMSEs of BEL, ANFIS, and MLP [34]. Table I also presents the NMSEs of peak values' prediction using BELPM and other data-driven models.

Figure 6 shows the percentage of peak values' identification of 34 maximum values that are obtained from AE values of 1987 using the BELPM, ANFIS, and W-KNN. Using BELPM, the percentage of missed points, the maximum values that have not been identified correctly, decreases to 9%. Note that the points that have not been identified with the delays or advance less than or equal to two days are considered as missed points.

TABLE I
A COMPARISON BETWEEN THE NMSE INDEX OF BELPM AND THE NMSE INDICES OF OTHER DATA-DRIVEN MODELS TO PREDICT AE INDEX OF 1978.

| Method | Learning algorithm | NMSE of PEAK point | NMSE of all points |
|---|---|---|---|
| BELPM | Emotional 45nodes for BL and MO | 0.3655 | 0.7989 |
| BEL[34] | Emotional 4reinforcmenr | 0.4411 | 0.971 |
| T_S[34] | 18 rules | 0.9883 | 0.9418 |
| MLP[34] | BP- 35neuron | 1.699 | 1.226 |

To show the preference of BELPM for its accurate prediction of the complex system using limited data, the BELPM is applied for one-day-ahead prediction of an AE time series using the training set and test set, which are selected from the same year (1987). First, the AE values of 6 months (180 days) are considered as training data to predict the AE time series of the next 6 months. Then the number of training samples is increased and the AE index of 304 days are considered as training pairs to forecast the AE values of the next sixty days. The bar charts of Fig. 7 show the NMSEs of peak values of the one day ahead prediction using different methods versus different training samples. The results not only show the ability of BELPM to learn from a limited training data set (180 samples), but they also indicate that decreasing the number of training samples does not make a noticeable difference between the NMSEs of BELPM; in contrast to ANFIS for which the NMSE increases with the decrease in the size of the training set. The obtained NMSEs from applying BELPM for these two cases are less than the NMSEs of ANFIS and W-KNN. Table II compares the NMSE, the specification and the predicted value of peak point by applying different methods that utilize the AE values of 304 days as training data.

TABLE II
A COMPARISON BETWEEN BELPM AND OTHER METHODS TO FORECAST THE AE INDEX USING DATA OF 300 DAYS.

| Method | Specification | NMSE of PEAK points | Predicted PEAK values |
|---|---|---|---|
| BELPM | 8 nodes for BL and MO | 0.0587 | 450.56 |
| ANFIS | 4rule | 0.1777 | 404.09 |
| WKNN | 2 neighbor | 0.0941 | 478.39 |

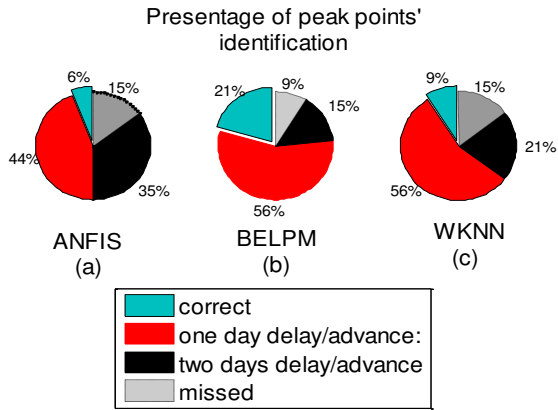

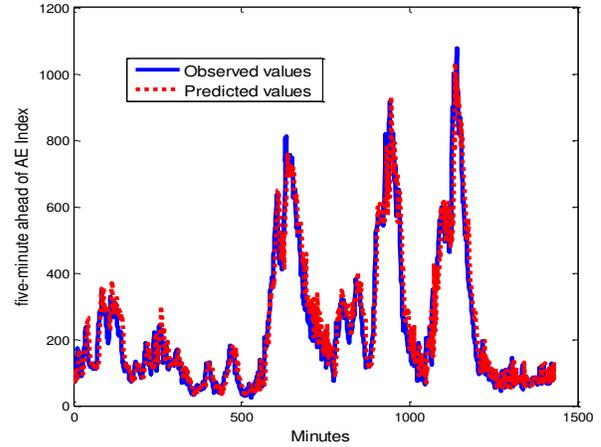

Fig.6. The pie charts of peak points' identification using different methods.

Fig.8. The predicted values of five- minute ahead of AE by BELPM.

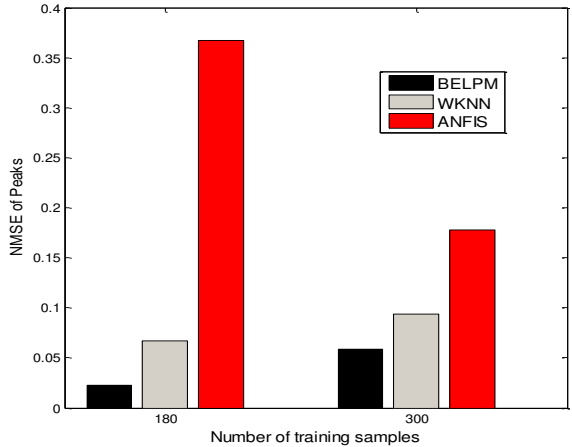

Fig. 7. The NMSE of peak points' prediction versus different training data set by BELPM, ANFIS and WKNN.

In the second experiment of this subscetion, BELPM is employed for short-term prediction of the AE index. For this purpose, the AE index of March 1992 is utilized as a prediction dataset. Different methods, such as the BELPM, ANFIS, and W-KNN are applied to the multi minute-ahead prediction of the AE values of the $9^{th}$ day. For the short term prediction, three training data sets are considered: the AE values of one day ($7^{th}$), two days ($6^{th}$ and $7^{th}$) and four days ($3^{rd}$ to $7^{th}$). Table III lists the NMSEs of different methods using the different training data sets for five-minute ahead prediction.

TABLE III
A COMPARISON BETWEEN BELPM AND OTHER METHODS TO FORECAST THE FIVE MINUTE AHEAD OF AE INDEX.

| Method | NMSE of 1Day ($7^{th}$) | NMSE of 2Days ($6^{th}$ and $7^{th}$) | NMSE of 4Days ($3^{rd}$ to $7^{th}$) |
|---|---|---|---|
| BELPM | 0.0802 | 0.0757 | 0.0776 |
| ANFIS | 0.0913 | 0.0784 | 0.0805 |
| WKNN | 0.0858 | 0.0834 | 0.0799 |

Figure 8 shows the five-minute ahead predictions of the $9^{th}$ day using BELPM as the prediction method and the AE values of the seventh day as the training samples. As another experiment, the BELPM, ANFIS, and WkNN are trained by the AE values of four days from $3^{rd}$ to $7^{th}$ day to predict the thirty minute ahead of the AE time series. Table IV compares the NMSEs, the CPU time, and the predicted values of peak point. Although the NMSE of BELPM is not low, this method predicts the peak value of the $9^{th}$ more accurately than others. Table IV also indicates that the computation time of BELPM, using the 5820 samples of four days, is much higher than the WKNN method. In fact, when there is a large number of a sample as training data, e.g. 5820, the computational time of the BELPM increases. However, the size of the training data set has not only an impact on time complexity of BELPM, it also raises the computational time of other methods such as ANFIS (see table IV). In comparison with other methods, the results of BELPM are more accurate, while its computation time has not a noticeable difference with the others except for WkNN. The predicted value of the $9^{th}$ day's peak values using BELPM is closer to the observed value (see table IV). Figure 9 depicts the NMSEs of ANFIS, WkNN and BELPM versus the prediction horizon. It can be seen that the NMSE of these methods have nearly the same values when the prediction horizon is lower than 20 minutes. The increase in NMSE is caused by the increase in the prediction horizon. However, the NMSE of BELPM is lower than WkNN and ANFIS, especially for higher prediction horizons (30 minute to 35 minute).

TABLE IV
A COMPARISON BETWEEN BELPM AND OTHER METHODS TO FORECAST THE THIRTY MINUTE AHEAD OF AE INDEX.

| Method | NMSE | Delay for the peak | CPU TIME (Minute) |
|---|---|---|---|
| BELPM | 0.4416 | No delay | 9.64 |
| ANFIS | 0.4473 | Two days delays | 9.70 |
| KNN | 0.4660 | One day advance | 0.1267 |

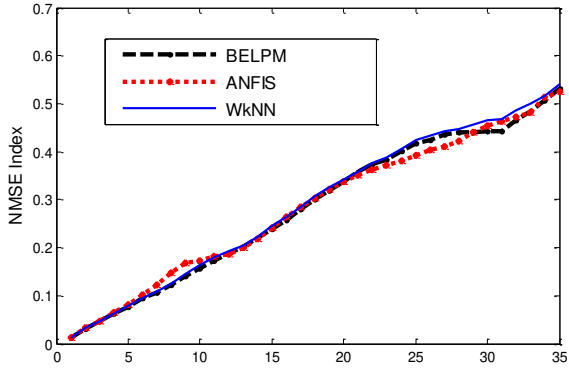

Fig.9. The NMSEs of multi-minute-ahead prediction of AE time series.

TABLE V
A COMPARISON BETWEEN BELPM AND OTHER METHODS TO PREDICTION OF AE INDEX.

| Method | Prediction Horizon | Year | Performance measurement |
|---|---|---|---|
| ANN[14] | One minute ahead | 1973-1974 | Correlation=0.86 |
| ANN[35] | One hour ahead | 2001,2006,2007 | Correlation=0.83 |
| BELPM | One minute ahead | 1992 | Correlation =0.99 |
| BELFIS | One minute ahead | 1992 | Correlation =0.99 |
| BELRFS | One minute ahead | 1992 | Correlation =0.98 |

As was mentioned, the AE time series is a well-known time series for examining data-driven models. Tabel V summarizes the performance of different methods on the AE index.

The results of applying the BELPM for long-term and short-term prediction of the AE time series show excellent performance of the model and verify that it can be a useful alert tool for geomagnetic storm prediction.

### B. Prediction Disturbance Storm Time

As mentioned, Dst is another important geomagnetic indices; it has been proposed to characterize the phases of geomagnetic storms i,e., the initial phase, main phase and recovery phase, that depend on the minimum value of Dst.

In this subsection, BELPM is tested to predict the Dst index and the results are compared with the other studies. In this case, the embedded dimension is selected as three. In the first experiment, BELPM is tested to predict the Dst index that is related to one of the harmful geomagnetic storms which occurred during solar cycle 22. It caused severe damage to Quebec's electricity power system. Figure 10 depicts the hourly DST index during 1998 to 1999. Figure 11 shows the obtained predicted values by using ANFIS, WKNN and BELPM. It is obvious that BELPM is more successful to predict the minimum value of the Dst index. The correlation between the obtained values and the real values has been presented in Fig. 12.

The next experiment is related to examining BELPM for two-step ahead prediction of the Dst index. For this purpose, the Dst values during ten days of 2000 is predicted. The main goal of this study is to predict the geomagnetic storms of july 2000. Figure 13 shows the prediction results.

As was mentioned, the Dst time series is a well-known time series for the prediction application. Tabel VI summarizes the performance of different methods on the Dst index.

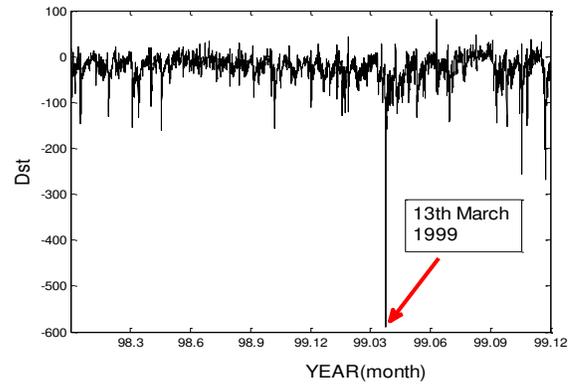

Fig.10. The Dst index of 1998 to 1999.

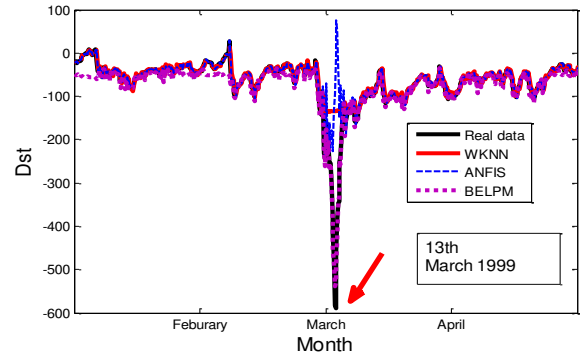

Fig.11. The predicted values versus the real values using different methods.

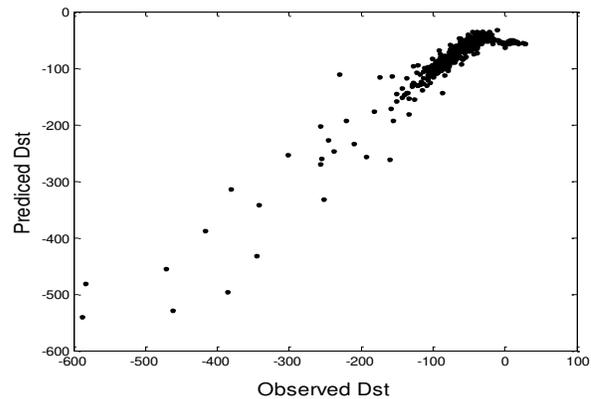

Fig.12. The correlation between the observed values and the predicted values.

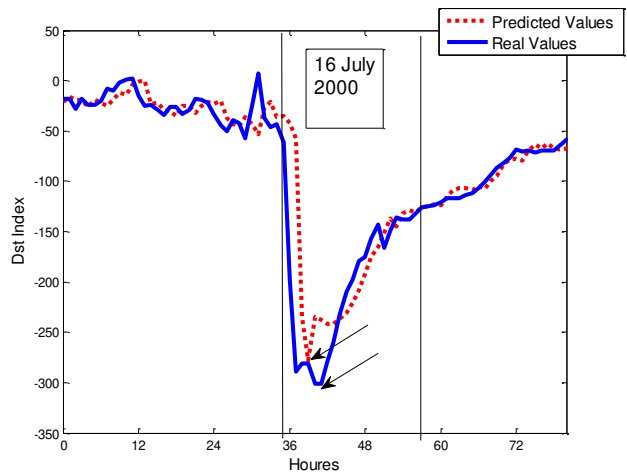

Fig.13. The predicted values of Dst index of 2000.

TABLE VI
A COMPARISON BETWEEN BELPM AND OTHER METHODS TO PREDICTION OF DST INDEX.

| Method | Prediction Horizon | Year | Performance measurement |
|---|---|---|---|
| ANN[13] | 1hour | 1998-1999 | Correlation=0.93 |
| ANN[13] | 6 hours | 1998-1999 | Correlation=0.75 |
| ANN[13] | 12 hours | 1998-1999 | Correlation=0.7 |
| ANN[13] | 18 hours | 1998-1999 | Correlation=0.65 |
| BELPM | 2 hours | 2000 | Correlation=0.94 |
| LoLiMoT[2] | 2 hours | 2000 | Correlation=0.94 |
| Anemomilos [15] | 24 hours | 2001 | Mean= 0.99 |
| Anemomilos[15] | 48 hours | 2001 | Mean= 0.91 |
| Anemomilos[15] | 72 hours | 2001 | Mean= 0.59 |
| Anemomilos [15] | 24 hours | 2005 | Mean= 0.77 |
| Anemomilos[15] | 48 hours | 2005 | Mean= 0.72 |
| Anemomilos[15] | 72 hours | 2005 | Mean= 0.58 |

## VI. CONCLUSION

This study has examined a prediction model inspired by brain emotional processing for geomagnetic storms. The accuracy of the BELPM has been extensively evaluated by different data sets of two benchmark indices of geomagnetic storms. The results strongly indicate that the model can be used for the long-term prediction of geomagnetic storms more accurately than other well-known methods, i.e. ANFIS. The results also show that BELPM is more efficient than the other methods when large training data sets are not available.

As future works, the author consider adding some optimization methods (e.g., genetic algorithm) to find optimal values of the fiddle parameters, e.g., the number of neighbors $k_a$ and $k_o$ and the initial values of nonlinear parameters. Other improvements in the model would be made on the basis of kd-Tree data structure [36] to address "the curse of dimensionality" [17] problem and decrease the computational time complexity of BELPM. To adjust the nonlinear parameters, different types of optimization methods (e.g., Quasi-Newton or Conjugate Directions) for ORBI and AMYG can be utilized. In addition, the Temporal Difference (TD) learning algorithm can also be used as a reinforcement method for the second learning phase to update the linear learning parameters. The good results obtained by employing the BELPM for predicting the chaotic time series are a motivation for applying this model as a classification method as well as to identify complex systems.